\documentclass[nonblindrev]{informs3}

\DoubleSpacedXII 

\usepackage{natbib}
\usepackage{caption}
\usepackage{subcaption}
\usepackage{threeparttable}
\usepackage{booktabs}
\usepackage{makecell}  
\usepackage{amsmath} 
\usepackage{algorithm}
\usepackage{algpseudocode}


\bibpunct[, ]{(}{)}{,}{a}{}{,}%
\def\bibfont{\small}%
\TheoremsNumberedThrough     

\EquationsNumberedThrough    

\begin{document}





\TITLE{Neural Jumps for Option Pricing}

\ARTICLEAUTHORS{%
\AUTHOR{Duosi Zheng\textsuperscript{1}, 
       Hanzhong Guo\textsuperscript{2}, 
       Yanchu Liu\textsuperscript{3*},
        Wei Huang\textsuperscript{1*}}

\AFF{\textsuperscript{1} College of Business, Southern University of Science and Technology, Shenzhen, China 
}
\AFF{\textsuperscript{2} School of Computing and Data Science, University of Hong Kong, Hong Kong, China 
}
\AFF{\textsuperscript{3} Lingnan College, Sun Yat-sen University, Guangzhou, China 
}
\AFF{\textsuperscript{*} Corresponding authors: liuych26@mail.sysu.edu.cn; huangw7@sustech.edu.cn.}

} 
\date{}
\ABSTRACT{
Recognizing the importance of jump risk in option pricing, we propose a neural jump stochastic differential equation model  in this paper, which integrates neural networks as parameter estimators in the conventional jump diffusion model. To overcome the problem that the backpropagation algorithm is not compatible with the jump process, we use the Gumbel-Softmax method to make the jump parameter gradient learnable. We examine the proposed model using both simulated data and S\&P 500 index options. 
The findings demonstrate that the incorporation of neural jump components substantially improves the accuracy of pricing compared to existing benchmark models. 
}


\maketitle 
\clearpage 


\section{Introduction}\label{intro}
\cite{black1973pricing} establish the fundamental framework for option pricing.
However, extensive empirical studies revealed that this seminal framework does not capture volatility smiles and leptokurtic in the return distribution
(\citeauthor{kou2002jump}, \citeyear{kou2002jump}; \citeauthor{kim2021portfolio}, \citeyear{kim2021portfolio}).
Consequently, various extensions have been developed (\citeauthor{cox1976valuation}, \citeyear{cox1976valuation}; \citeauthor{dupire1994pricing}, \citeyear{dupire1994pricing}; \citeauthor{hull1987pricing}, \citeyear{hull1987pricing}; \citeauthor{heston1993closed}, \citeyear{heston1993closed}). 
Among them, jump models have received particular attention due to their ability to explain abrupt and discontinuous movements in asset returns and volatilities (\citeauthor{merton1976option}, \citeyear{merton1976option}; \citeauthor{bates1996jumps}, \citeyear{bates1996jumps}; \citeauthor{duffie2000transform}, \citeyear{duffie2000transform}).
Empirical investigations have also shown that the jump component is an important element in pricing options (\citeauthor{eraker2003impact}, \citeyear{eraker2003impact}; \citeauthor{cummins2025appraising}, \citeyear{cummins2025appraising}).
Despite these advancements, parametric extensions remain constrained by inherent structural assumptions, which may not fully accommodate complex financial markets (\citeauthor{bates2003empirical}, \citeyear{bates2003empirical}).

In contrast, nonparametric methods driven by data directly approximate pricing functions without restrictions. Since the pioneering contribution of \cite{hutchinson1994nonparametric} demonstrates the efficacy of artificial neural networks (ANN) in capturing the dynamics of option prices and implementing hedging strategies, it has become one of the most influential data-driven methods (\citeauthor{amilon2003neural}, \citeyear{amilon2003neural};  \citeauthor{liu2019pricing}, \citeyear{liu2019pricing}). 
\cite{ruf2020neural} present an early review on applications of ANN in option pricing and hedging.
However, the ANN method does not have a sounding theory that supports the training process. In addition, to obtain a well-trained network, a large-scale data set is generally required.

The emergence of hybrid models that integrate deep learning with parametric option pricing models has recently gained research momentum.
\cite{andreou2008pricing} propose a hybrid network that incorporates information from Black and Scholes (BS) implied volatilities.
\cite{cao2021option} construct a neural network architecture that enforces no-arbitrage conditions in the selection of input layer weights.
\cite{das2017new} investigate a model using homogeneity hints to group the option price estimated by parametric models, and feed these groups of estimated price into neural networks. 
\cite{shvimer2024pricing} propose a two-submodel framework with moneyness-adaptive parameterization.
These studies extract additional information from parametric models and incorporate it into neural networks to enhance empirical accuracy, but do not establish a foundational connection between neural networks and parametric models.

\cite{weinan2017proposal} establishes a theoretical bridge between neural networks and discrete dynamical systems, suggesting that deep learning architectures can be interpreted as discretized forms of differential equations. 
\cite{chen2018neural} introduce neural ordinary differential equations (NODE) that model the derivative function of ordinary differential equations through neural networks.
Researchers have extended the framework of neural differential equations to other dynamical systems (\citeauthor{kidger2020neural}, \citeyear{kidger2020neural}; \citeauthor{li2020scalable}, \citeyear{li2020scalable}; \citeauthor{khoo2021solving}, \citeyear{khoo2021solving}).
Following similar insights, \cite{wang2021option} propose a neural-diffusion stochastic differential equation (NSDE) model for option pricing, where deterministic parameters are modeled as nonlinear functions. This allows the drift and volatility terms to vary dynamically over time, mitigating model misspecification and enhancing adaptability.
\cite{maneural} combine neural networks with a rough volatility model, further improving the modeling accuracy of volatility dynamics. 
\cite{halskov2023deep} constructs a deep structural model employing neural networks to estimate conditional firm-level parameters in the Merton-type model.
Despite these innovations retaining stochastic structures in parametric models, they predominantly concentrate on continuous-time models and neglect the jump process, which has been empirically identified as a crucial factor in option pricing (\citeauthor{eraker2003impact}, \citeyear{eraker2003impact}; \citeauthor{cummins2025appraising}, \citeyear{cummins2025appraising}).


This study develops a novel neural jump stochastic differential equation (NJSDE) model. Aiming at solving the incompatibility of the backpropagation (BP) algorithm with jump process, and enabling the whole training process to be continuously differentiable.
Our paper is close to \cite{chen2023deep} which also develops a surrogate model for jump-diffusion models. They define model parameters as pseudo-state variables and randomly sample them within a predefined empirical range as input of a neural network. This approach reconstructs a mapping between parameters and predicted prices, thus avoiding the curse of dimensionality associated with complex structural models. However, this newly constructed relationship omits the original model structure, which is based on theoretical assumptions.
Our approach in this paper retains the stochastic terms from the parametric model to preserve fundamental economic assumptions, while replacing the deterministic components with neural networks to enhance model flexibility.

Incorporating neural networks into structural models renders analytical solutions intractable.  
Following \cite{wang2021option}, we reformulate the discretized structural model as a recurrent neural network-like structure and train it using BP algorithm.
\cite{jia2019neural} extend the NODE model and propose a likelihood-based approach to optimize the jump process. This method enables scalable gradient computation via the adjoint sensitivity method rather than the BP algorithm, thereby avoiding the incompatibility of the BP algorithm with the jump process. However, this approach may introduce additional computational complexity and potential truncation errors (\citeauthor{ma2021comparison}, \citeyear{ma2021comparison}).
Compared to \cite{jia2019neural}, our model is developed in the context of option pricing and specifically focuses on handling jump process with the BP algorithm.

It's well known that jump diffusion models exhibit discontinuities, making gradient-based deep learning methods less effective.
Furthermore, the randomness of the jump process is intrinsically governed by the jump intensity parameter, making it difficult to separate the random term from the jump intensity parameter through reparameterization methods. 
To address the nondifferentiability of the jump parameters, we adopt the Gumbel-Softmax method (\citeauthor{jang2017categorical}, \citeyear{jang2017categorical}), which enables a differentiable relaxation of the jumps, allowing gradient-based optimization within the BP algorithm.
Theoretically, we can utilize the powerful ability of neural networks to approximate any function, thereby approximating the coefficient function in the parametric model (\citeauthor{hornik1989multilayer}, \citeyear{hornik1989multilayer}). 
Empirical tests demonstrate that the proposed NJSDE model achieves the lowest pricing error in the presence of jumps, effectively capturing discontinuous dynamics. Even in the absence of jumps, the NJSDE model maintains a competitive performance comparable to the NSDE model, showing its robustness under different market conditions.

Our methodology offers at least two primary contributions to the literature.
First, we introduce a novel option pricing model that systematically integrates the jump process with deep learning methods. 
Our approach addresses the challenge of the incompatibility between the jump process and gradient-based neural network optimization. To our best knowledge, this is the first attempt to optimize the jump process through the BP algorithm in the option pricing domain.
Our methodology bridges the gap between jump models and neural networks, enabling a more effective calibration of jump risk in financial markets. 

Second, the proposed model presents a more general hybrid structure combining the strengths of both neural networks and jump models. 
The selection of the option pricing model involves a choice among misspecified models. Our model is sufficiently flexible to encompass most widely used parametric models, treating them as special cases within the broader framework.
In particular, depending on market conditions, the NJSDE model can degenerate into stochastic volatility models or non-jump models, making it a unified and adaptive framework across different financial conditions.

The remainder of this paper is organized as follows.  
Section \ref{model} introduces the proposed methodology. Section \ref{Synthetic Data} and Section \ref{Real Data} employ synthetic and real market data, respectively, to showcase the empirical performance of our approach. Section \ref{conclusion} concludes the paper.

\section{Model and Estimation}\label{model}
This section first introduces the Gumbel-Softmax method, which we use to reparameterize the deterministic terms of the jump process, followed by the full specification of the NJSDE model. We then present how to calibrate model parameters and estimate option prices.

\subsection{The Gumbel-Softmax Method}
Let the occurrence of jumps satisfy a Poisson process $N_t$ with intensity parameter $\lambda$. To enable gradient-based optimization through the discrete jump structure, we employ the Gumbel-Softmax method (\citeauthor{jang2017categorical}, \citeyear{jang2017categorical}), which provides a continuous and differentiable approximation to sample from a categorical distribution, while preserving the original probability structure.

Let $\pi_i$ denote the normalized probability of observing $i$ jumps within a given time $t$, $i= 0,1,2 \dots, n$, where $n$ is a user-specified upper bound on the number of jumps. This truncation assumes that at most $n$ jumps can occur within $t$, and $n$ acts as a hyperparameter in the model. 
The probability $\pi_i$ is given by:
\begin{equation}
\pi_i = \frac{P(N_t = i)}{\sum_{j=0}^n P(N_t = j)}
= \frac{\frac{(\lambda t)^i}{i!}e^{-\lambda t}}{\sum_{j=0}^n \frac{(\lambda t)^j}{j!}e^{-\lambda t}}
= \frac{\frac{(\lambda t)^i}{i!}}{\sum_{j=0}^n \frac{(\lambda t)^j}{j!}}.  \label{equ_1}
\end{equation}


We use $z$ to denote the realized number of jumps occurring within $t$, and the categorical sampling process can be first expressed using the Gumbel-Max method (\citeauthor{gumbel1954statistical}, \citeyear{gumbel1954statistical}):
\begin{equation}
    z = \mathop{\arg\max}\limits_{i} (g_i + \log \pi_i) \label{equ_2},  
\end{equation}
where $g_i \in \{g_0, g_1, \dots, g_n\}$ is a random variable sampled independently from a standard Gumbel distribution. The inclusion of $g_i$ allows the inherently discrete sampling of jumps to be reformulated as a differentiable transformation, effectively shifting the randomness from the jump process to the parameter-free Gumbel distribution.

To obtain a differentiable approximation to the non-differentiable $\arg\max$ function in Equation~\eqref{equ_2}, we apply the softmax function:
\begin{equation}
    y_i = \mathop{softmax}\limits_{i} (g_i + \log \pi_i) = \frac{\exp((g_i + \log \pi_i) / \tau)}{\sum_{j=0}^n \exp((g_j + \log \pi_j) / \tau)}
    \label{equ_3},
\end{equation}
where $y_i$, for $i= 0,1,2 \dots, n$, denotes the relaxed probability (via Gumbel-Softmax method) of observing $i$ jumps. $\tau > 0$ is a temperature parameter controlling the degree of approximation. As $\tau \to 0$, the Gumbel-Softmax samples approach one-hot, recovering the original categorical samples.

\subsection{Neural Jump Stochastic Differential Equation (NJSDE) Model}
We start from the stochastic volatility with correlated jumps (SVCJ) model in \cite{duffie2000transform}. The SVCJ model extends classical stochastic volatility models by incorporating simultaneous jumps in both the asset price and its volatility process, thereby capturing more realistic dynamics observed in financial markets. 
The specification of the SVCJ model is given below:
\begin{equation}
    d\log S_t = \mu dt + \sqrt{V_t}dW_{t}^{(S)} + Z_{t}^{y} dN_t 
    \label{equ_4},
\end{equation}
\begin{equation}
    dV_t = \kappa (\theta -V_t)dt + \sigma_V \sqrt{V_t}dW_{t}^{(V)} + Z_{t}^{v} dN_t 
    \label{equ_5},
\end{equation}
where $S_t$ and $V_t$ represent the asset price and volatility process at time $t$, respectively. 
$W_{t}^{(S)}$ and $W_{t}^{(V)}$ are standard Brownian motions with a correlation coefficient $\rho$. 
$N_t$ denotes a Poisson process with constant intensity $\lambda$. 
$Z_t^y$ and $Z_t^v$ are jump sizes in the asset price and volatility process, respectively.

We retain stochastic terms in Equations~\eqref{equ_4} and \eqref{equ_5}, while the deterministic terms originally defined by specific assumptions, including but not limited to $\mu, \kappa$ and $\theta$, are replaced by neural network components $NN_i$, for $i = 1, \dots, l$, where $l$ depends on how many model parameters need to be approximated by neural networks. 
Each $NN_i$ represents a feedforward neural network served as a nonparametric estimator of the corresponding deterministic term. All networks share the same input features, including the asset price $S_t$, strike price $K$, time to maturity $t$, and risk-free rate $r_f$. These networks are independently parameterized with different weights to ensure flexible learning of distinct functional relationships. 
By substituting predefined parameters with learnable neural networks, the proposed model alleviates the need for strong structural assumptions. Instead, model dynamics are inferred directly from data, enabling greater adaptability in capturing complex financial behavior.
The main NJSDE model expressions are given below:
\begin{equation}
    dS_t = NN_1 dt + NN_2 dW_t^{(S)} + NN_3 U_t^{(S)} f(NN_7) ,\label{equ_7}
\end{equation}
\begin{equation}
    dV_t = NN_4 dt + NN_5 dW_t^{(V)} + NN_6 U_t^{(V)} f(NN_7), \label{equ_8}
\end{equation}
where $NN_1, NN_2, NN_4$, and $NN_5$ replace the drift and diffusion terms in the asset price and volatility process directly. Jump sizes $Z_t^y$ and $Z_t^v$ are assumed to follow continuous distributions, we apply the reparameterization method to express them as the product of two components: a deterministic term parameterized by $NN_3$ and $NN_6$, respectively, and a stochastic term $U_t^{(S)}$ and $U_t^{(V)}$, which are drawn from the uniform distribution on (0, 1), respectively.

We approximate the jump term in Equations~\eqref{equ_4} and \eqref{equ_5} using the Gumbel-Softmax method, which is $f(NN_7)$ in Equations~\eqref{equ_7} and \eqref{equ_8}, and the jump intensity $\lambda$ is replaced by a neural network estimator $NN_7$. Applying Equation~\eqref{equ_1} within the time increment $dt$, the probability of observing $i$ jumps is:
\begin{equation}
    \pi_i = \frac{(NN_7 dt)^i / i!}{\sum_{j=0}^n (NN_7 dt)^j / j!}. \label{equ_9}
\end{equation}

The correlation $\rho$ between Brownian motions $W_{t}^{(S)}$ and $W_{t}^{(V)}$ is also adaptively learned by $NN_{8}$. Specifically, we define $W_{t}^{(V)}$ as follow, where $W_t$ is an independent Brownian motion:
\begin{equation}
    dW_t^{(V)}= NN_{8} dW_t^{(S)} + \sqrt{1-(NN_{8})^2}dW_t.
    \label{equ_NN_rho}
\end{equation}


\subsection{Model Calibration}
Since neural networks are embedded within the jump diffusion model, analytical solutions are not available. Therefore, we employ the Monte Carlo method to numerically approximate the option price.  
Taking European call option as an example, the option price can be expressed as:
\begin{equation}
    \overline P = \frac{1}{M} \sum_{k=1}^{M} \left\{e^{-r_f T}(S_T^k - K)^+\right\}, \label{equ_13}
\end{equation}
where $\overline P$ is the estimated option price, $S_T^k$ is the asset price in the $kth$ simulation path at expiration date $T$ for $k = 1, 2, \dots, M$, and $x^+:=max\{x,0\}$.

The calibration of the model is performed using Euler discretization. The discretized dynamics of the asset price $S_t$ and volatility $V_t$ are generate at $m$ discrete time points within $T$, where $0 \leq t \leq T$ and $\Delta t = T /m$:
\begin{equation}
    S_{t+\Delta t} = S_t + NN_1 \Delta t + NN_2 \sqrt{\Delta t} \varepsilon_t^{(S)} + NN_3 U_t^{(S)} f(NN_7), \label{equ_10}
\end{equation}
\begin{equation}
    V_{t+\Delta t} = V_t + NN_4 \Delta t + NN_5 \sqrt{\Delta t} \varepsilon_t^{(V)} + NN_6 U_t^{(V)} f(NN_7), \label{equ_11}
\end{equation}
where $\varepsilon_t^{(S)}$ and $\varepsilon_t^{(V)}$ are two random variables following a standard Normal distribution.
The model after discretization can be treated as a recursive expression, analogous to a specialized form of recurrent neural networks, which allows for efficient training using BP algorithm.

\begin{figure}
    \caption{Structure of Neural Jump Stochastic Differential Equation (NJSDE) Model} 
    \label{fig:structure}
    \includegraphics[width=\textwidth]{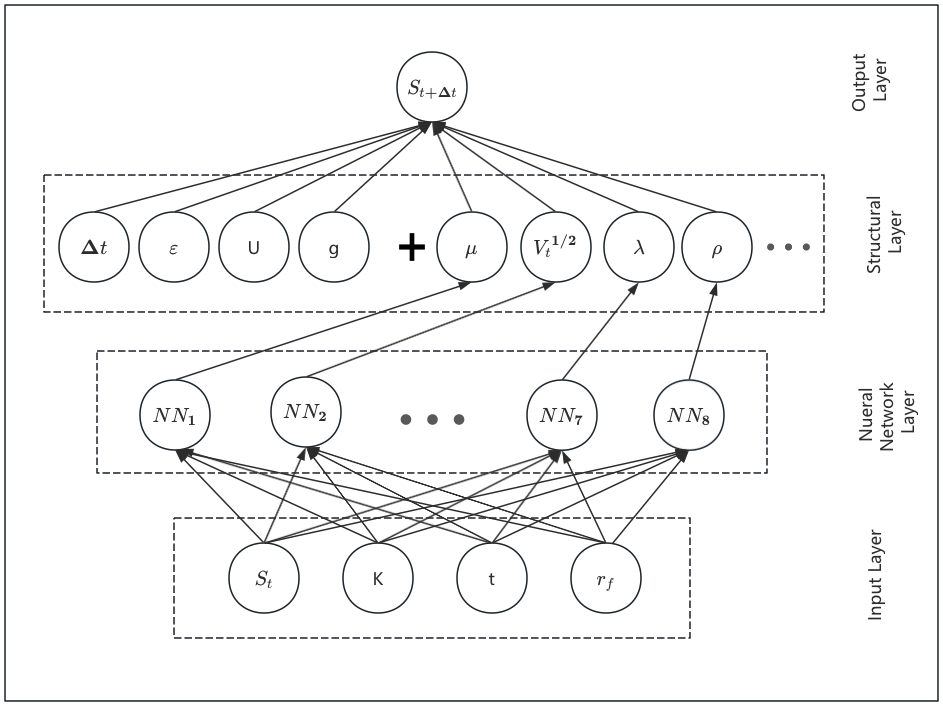}
    
    \smallskip
    \footnotesize\textit{Notes.} \cite{halskov2023deep} has similar plots in spirit but under different settings.
\end{figure}

A visualization of the model structure is presented in Figure~\ref{fig:structure}. The input layer consists of four features: $S_t$, $K$, $t$, $r_f$. These features are fed into eight independently parameterized neural networks $\mathrm{NN}_i$, for $i = 1, 2, \dots, 8$, which together constitute the neural network layer. Combining neural networks with stochastic terms and $\Delta t$ in the structural layer, we then get $S_{t+\Delta t}$ in the output layer. 

Let $\omega_i$ be the vector of all trainable parameters of $\mathrm{NN}_i$, for $i = 1, 2, \dots, 8$, and define the full parameter set as $\omega = (\omega_1, \omega_2, \omega_3, \omega_4, \omega_5, \omega_6, \omega_7, \omega_8)$.
The loss function $L$ for calibration is defined as:
\begin{equation}
    L = \sum_{j=1}^{I}[P_j - P_j(\omega)]^2, \label{equ_14}
\end{equation}
where $I$ represents the total number of options with different strikes and maturities. $P_j$ is the target price and $P_j(\omega)$ is the estimated price of the $jth$ option.

The calibration task is then formulated as the following optimization problem:
\begin{equation}
    \min_{\omega} \sum_{j=1}^{I} [P_j - P_j(\omega)]^2. \label{equ_minimize}
\end{equation}


Taking $\omega_1$, the trainable parameters of $NN_1$, as an example, we compute the gradient of loss function using the chain rule:
\begin{equation}
    \frac{\partial L}{\partial \omega_1} = \frac{\partial L}{\partial P_j(\omega)} \frac{\partial P_j(\omega)}{\partial \omega_1} = 2\sum_{j=1}^{I}[P_j - P_j(\omega)] \frac{\partial P_j(\omega)}{\partial \omega_1}. \label{equ_15}
\end{equation}

Applying Equation~\eqref{equ_13}, the gradient of the estimated price with respect to $\omega_1$ can be expressed as:
\begin{equation}
    \frac{\partial P_j(\omega)}{\partial \omega_1} = e^{-r_f T_j}\frac{1}{M} \sum_{k=1}^{M} [ \frac{\partial S_{T_j}^k}{\partial \omega_1}\mathbb{I}_{\{S_{T_j}^k-K_j>0\}}], \label{equ_16}
\end{equation}
where $T_j$ and $K_j$, $j = 1, 2,\dots , I$, are the maturity and strike price of the $jth$ option, respectively, and $\mathbb{I}_{\{\cdot\}}$ is the indicator function. The recursive gradient of $S_{t+\Delta t}^k$ with respect to $\omega_1$ is given as below, where $0 \leq t \leq T_j$ and $\Delta t = T_j / m$ for the $jth$ option:
\begin{equation}
    \begin{split}
        \frac{\partial S_{t+\Delta t}^k}{\partial\omega_1} = \frac{\partial NN_1}{\partial\omega_1} \Delta t + \frac{\partial S_{t}^k}{\partial\omega_1} [1+\frac{\partial NN_1}{\partial S_{t}^k} \Delta t 
         + \frac{\partial NN_2}{\partial S_{t}^k} \sqrt{\Delta t} \epsilon_{t}^{(S)}
         \\ +\frac{\partial NN_3}{\partial S_{t}^k} U_{t}^{(S)} f(NN_7)+NN_3 U_{t}^{(S)} \frac{\partial f(NN_7)}{\partial S_{t}^k}]. \label{equ_17}
    \end{split} 
\end{equation}

These gradients can be efficiently computed using the BP algorithm, taking advantage of the differentiable structure of the neural network and the Monte Carlo simulation framework. According to the chain rule, the gradients with respect to other parameters $\omega_i$, for $i = 1, 2, \dots, 8$, can be obtained in the same way.
The main simulation steps are summarized in Algorithm~\ref{alg:algorithm_1}.
\begin{algorithm}[h]
\small 
\caption{Simulation Procedure for the NJSDE Model}
\label{alg:algorithm_1}
\begin{algorithmic}[1]  
    \State Construct eight neural network structures $NN_i$, for $i=1,2, \dots, 8$, with appropriate activation functions and numbers of layers.
    \State Set the time step size $\Delta t = T/m$ in Equation~\eqref{equ_10}, where $T$ represents the expiration date and $m$ indicates the total number of time steps.
    \State Specify the number of training epochs $D$ and the number of Monte Carlo sample paths $M$ per epoch.
    \State Generate $m \times M$ standard random variables for $\varepsilon_t^{(S)}, \varepsilon_t^{(V)}, U_t^{(S)}$ and $U_t^{(V)}$.
    \State Initialize $S_0$ and $V_0$. Simulate $M$ sample paths of $S_t$ and $V_t$ using Equations~\eqref{equ_10} and \eqref{equ_11}, and compute option prices via Equation~\eqref{equ_13}.
    \State Minimize the loss function defined in Equation~\eqref{equ_14}, and update the neural network parameter set $\omega$ using BP algorithm. Repeat from Step 5 until the number of epochs reaches $D$.
\end{algorithmic}
\end{algorithm}

\section{Numerical Experiments} \label{Synthetic Data}
In this section, we investigate the predictive performance of the proposed model using simulated data. 
Specifically, we assume that the simulated data follows either the Heston model (\citeauthor{heston1993closed}, \citeyear{heston1993closed}) or the SVCJ model (\citeauthor{duffie2000transform}, \citeyear{duffie2000transform}). 

We compare the performance of the NJSDE model with three parametric models: the BS model (\citeauthor{black1973pricing}, \citeyear{black1973pricing}), the Heston model and the SVCJ model. Additionally, we consider an ANN model, which is a classical nonparametric method widely adopted in the literature, with over 150 papers applying it to option pricing and hedging (\citeauthor{ruf2022hedging}, \citeyear{ruf2022hedging}), and the neural stochastic differentiable equation (NSDE) model (\citeauthor{wang2021option}, \citeyear{wang2021option}).

Model accuracy is evaluated using two standard statistical indicators: mean absolute error (MAE) and mean squared error (MSE), both of which quantify the deviation between actual and predicted option prices (\citeauthor{andreou2010generalized}, \citeyear{andreou2010generalized}; \citeauthor{shvimer2024pricing}, \citeyear{shvimer2024pricing}). These metrics are defined as follows:
\begin{equation}
    MAE=\frac{1}{I} \sum_{j=1}^I\left|C_{j, actual}-C_{j, forecast}\right|,
\end{equation}
\begin{equation}
    MSE=\frac{1}{I} \sum_{j=1}^I\left(C_{j, actual}-C_{j, forecast}\right)^2,
\end{equation}
where $C_{j, actual}$ is the actual option price and $C_{j, forecast}$ is the forecast option price of the $jth$ option, and $I$ represents the total number of options.


\subsection{Heston model}\label{experiment 2}
The Heston model assumes that the asset price $S_t$ and volatility $V_t$ follow the dynamics:
\begin{equation}
    dS_t = \mu S_t dt + \sqrt{V_t} S_t dB_{1, t}, 
\end{equation}
\begin{equation}
    dV_t = \kappa (\theta -V_t)dt + \sigma \sqrt{V_t}dB_{2, t}, 
\end{equation}
where $B_{1, t}$ and $B_{2, t}$ are two correlated standard Brownian motions with the correlation coefficient $\rho$. 
The parameters utilized to generate the numerical samples are specified as follows: $\mu$ = 0.04, $\kappa$ = 1.5, $\theta$ = 0.1, $\sigma$ = 0.3, $\rho$ = -0.5. The initial values $S_0$ and $V_0$ are set at 100 and 0.04, respectively, and the risk-free rate $r_f$ is 0.025. 
In the path simulation, the time to maturity $T$ and the strike price $K$ in the training set are determined by $[\frac{1}{12}, \frac{2}{12}, \frac{3}{12}, \frac{6}{12}, 1]$ and $[60, 70, 80, 90, 100, 110, 120, 130, 140]$, respectively. For the testing set, $T$ and $K$ are determined by $[\frac{1}{12}, \frac{2}{12}, \frac{3}{12}, \frac{4}{12}, \frac{5}{12}, \frac{6}{12}, \frac{8}{12}, \frac{9}{12}, \frac{10}{12}, 1] $ and $ [60, 65, 70, 75, 80, 85, 90, 95, 100, 105, 110, 115, 120, 125, 130, 135, 140]$, respectively.
This setup ensures broad coverage across different levels of moneyness and maturities, and each sample point corresponds to a unique $(T, K)$ pair under the specified parameter settings.
The parameter values and the sample point settings follow the experimental setup in \cite{wang2021option}.

\begin{table}[h]
    \setlength{\tabcolsep}{8pt}  
    \caption{\textbf{Overall Pricing Performance of Synthetic Options Generated from the Heston Model}}
    \label{tab:Heston_pricing}
    \centering
    \small  
    \renewcommand{\arraystretch}{0.8} 
    \begin{tabular}{llccccc}
        \toprule
        & & BS & Heston & ANN & NSDE & NJSDE \\
        \midrule
        In-sample & MAE & 0.4939 & 0.3817 & 0.9479 & 0.2280 & 0.2635 \\
                  & MSE & 1.1126 & 0.4891 & 2.3551 & 0.1123 & 0.1595 \\
        \midrule
        Out-of-sample & MAE & 0.6077 & 0.4710 & 0.6430 & 0.2409 & 0.2519 \\
                      & MSE & 1.3746 & 0.6417 & 1.0507 & 0.1666 & 0.1387 \\
        \bottomrule
    \end{tabular}
    \vspace{0.5em}
    
    \raggedright
    \footnotesize{\textit{Notes.} This table summarizes the in-sample and out-of-sample forecasting performance for synthetic options generated from the Heston model, using mean absolute error (MAE) and mean squared error (MSE).} 
\end{table}

Table \ref{tab:Heston_pricing} presents the overall fitting results. In total, compared to traditional parametric and nonparametric models, both the NJSDE and the NSDE models demonstrate improved predictive performance.
Specifically, the error between the NJSDE model and the NSDE model is minimal.
The NSDE model achieves the lowest in-sample errors, with an MAE of 0.2280 and an MSE of 0.1123. The NJSDE model reports slightly higher in-sample errors, with an MAE of 0.2635 and an MSE of 0.1595, but the difference remains small, suggesting that both models are highly effective in capturing the dynamics of the simulated data. In out-of-sample performance, the NSDE model continues to outperform slightly, attaining an MAE of 0.2409. However, the NJSDE model achieves the lowest MSE of 0.1387, indicating a potential benefit in minimizing squared errors.

These results suggest that the NSDE model can be interpreted as a specific case of the NJSDE model. 
When the jump components are not significant, the coefficients of the jump process, estimated via neural networks, tend to converge to zero, effectively simplifying the NJSDE model into the NSDE formulation. 
The slight discrepancy in performance may be attributed to the increased complexity of the NJSDE model, which can introduce minor estimation errors.

Furthermore, the ANN model exhibits significantly higher pricing errors compared to parametric models, likely due to its reliance on large datasets for effective training. Given the limited size of the simulated dataset, its underperformance is expected. 
In contrast, the NSDE and NJSDE models benefit from the structured parametric framework, which enhances model stability and enables accurate pricing even with moderate amounts of data.

\begin{figure}[h]
    \caption{\textbf{Pricing Performances of the Competing Models across Different Moneyness and Maturities}}
    \label{fig:Errors in experiment2}
    \begin{subfigure}{0.495\textwidth}
        \centering
        \includegraphics[width=\textwidth]{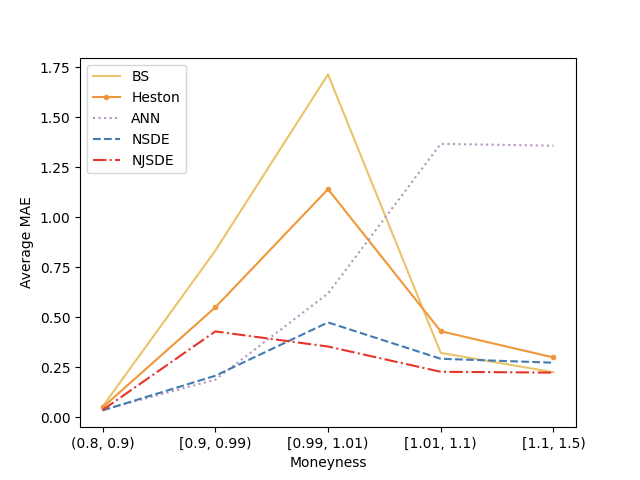}
        \caption{\small Out-of-Sample: All Maturities}
        \label{fig:E2_all}
    \end{subfigure}
    \hfill
    \begin{subfigure}{0.495\textwidth}
        \centering
        \includegraphics[width=\textwidth]{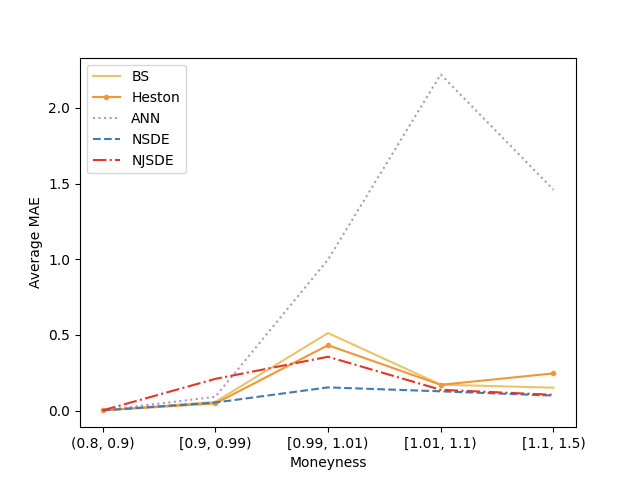}
        \caption{\small Out-of-Sample: Short-Term Maturity}
        \label{fig:E2_short}
    \end{subfigure}
    \hfill
    \begin{subfigure}{0.495\textwidth}
        \centering
        \includegraphics[width=\textwidth]{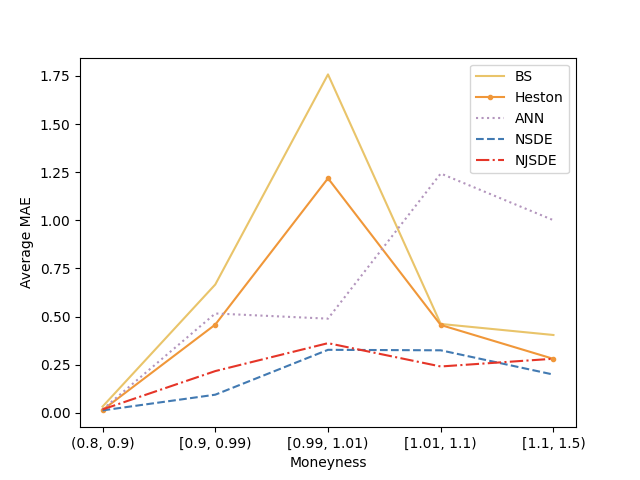}
        \caption{\small Out-of-Sample: Medium-Term Maturity}
        \label{fig:E2_medium}
    \end{subfigure}
    \hfill
    \begin{subfigure}{0.495\textwidth}
        \centering
        \includegraphics[width=\textwidth]{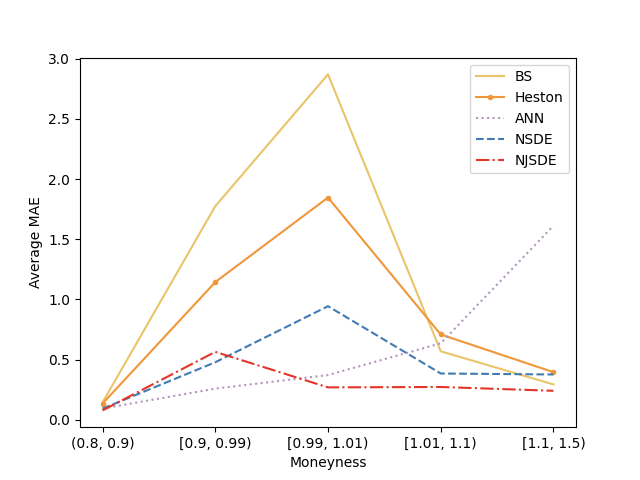}
        \caption{\small Out-of-Sample: Long-Term Maturity}
        \label{fig:E2_long}
    \end{subfigure}

    \smallskip
    \footnotesize\textit{Notes.} This figure shows the model’s out-of-sample performances across different moneyness and days to maturity brackets on the synthetic options generated from the Heston model, as measured by averages MAEs.
\end{figure}

To comprehensively assess out-of-sample performance, we visualize the distribution of average MAE across different days to maturity ($DTM$) and moneyness ($S/K$) levels.
Following \cite{ludwig2015robust}, options are categorized into five moneyness intervals: deep out-the-money (OTM) ($0.8<S/K<0.9$), OTM ($0.9\leq S/K<0.99$), at-the-money (ATM) ($0.99\leq S/K<1.01$), in-the-money (ITM) ($1.01\leq S/K<1.1$) and deep ITM ($1.1\leq S/K<1.5$).
$DTM$ are grouped into short-term (1 $\leq DTM <$ 60 days), medium-term (60 $\leq DTM <$ 180 days), and long-term ($DTM$ $\geq$ 180 days) maturity contracts.

Figure \ref{fig:E2_all} illustrates the out-of-sample pricing performance across whole maturities, while Figures \ref{fig:E2_short}–\ref{fig:E2_long} focus on short, medium and long-term maturity, respectively. The x-axis denotes moneyness categories, and the y-axis represents the corresponding average MAE values.
In total, the NSDE and NJSDE models consistently outperform other benchmark models. In particular, they exhibit stable performance across all the moneyness and maturity groups.
As shown in Figure \ref{fig:E2_short}, errors remain relatively low for short-term maturities across models except for the ANN model. However, longer maturities (Figures \ref{fig:E2_medium} and \ref{fig:E2_long}) highlight the increasing advantage of the NSDE and NJSDE models.

\begin{table}[h] 
    \setlength{\tabcolsep}{8pt}  
    \caption{\textbf{DM Test of Different Models for Synthetic Options Generated from the Heston Model}}
    \label{tab:DM_2}
    \centering
    \small  
    \renewcommand{\arraystretch}{0.8} 
    \begin{tabular}{ c c c c c }
        \toprule
        Model & Heston & ANN & NSDE & NJSDE \\
        \midrule
        BS & 
        3.69 (p$<$0.01) & 0.57 (p$=$0.56) & 6.72 (p$<$0.01) & 7.50 (p$<$0.01) \\
        Heston & - &
        1.87 (p$=$0.06) & 5.00 (p$<$0.01) & 4.21 (p$<$0.01) \\
        ANN & - & - &
        5.03 (p$<$0.01) & 5.31 (p$<$0.01) \\
        NSDE & - & - & - &
        0.65 (p$=$0.51) \\
        \bottomrule
    \end{tabular}
    \vspace{0.5em}
    
    \raggedright
    \footnotesize{\textit{Notes.} This table presents the DM values and p‐values for pairwise comparisons. A positive DM statistic indicates a preference for the column model over the row model.} 
\end{table}

The \cite{diebold2002comparing} (DM) test is employed to statistically compare the predictive performance of each pair of competing models, with results summarized in Table \ref{tab:DM_2}. 
Overall, the NJSDE model significantly outperforms the other benchmark models (p $<$ 0.01 in all relevant pairwise tests), except for the NSDE model.
The DM statistic between the NSDE and NJSDE models is 0.65 with a p-value of 0.51, suggesting no statistically significant difference in predictive performance. This result is consistent with the interpretation that the NJSDE model approximates the NSDE model in the absence of significant jump components.
Additionally, the ANN model does not exhibit statistically significant performance differences when compared with the BS (p = 0.56) and Heston (p = 0.06) models, likely due to limited training data.

\subsection{SVCJ model}\label{experiment 1}
The SVCJ model assumes that the asset price $S_t$ and volatility $V_t$ follow the dynamics specified in Equations~\eqref{equ_4} and \eqref{equ_5}.
To examine the ability of the NJSDE model to capture jump dynamics, we generate simulated option data incorporating jumps based on the SVCJ model. The parameter settings from the Heston model are retained and additional jump-related parameters in the SVCJ model are introduced. Specifically, these parameters are set as follows: $\lambda = 0.1, \mu_v = 0.6, \mu_y = 0.08, \sigma_y = 2.15, \rho_j = 0.57$.
The training and testing data point settings keep identical to those in the previous experiment. The NJSDE model is compared against the same benchmark models as before, including the SVCJ model, which serves as the underlying parametric framework for the NJSDE model. 

\begin{table}[h]
    \setlength{\tabcolsep}{8pt}  
    \caption{\textbf{Overall Pricing Performance of Synthetic Options Generated from the SVCJ Model}}
    \label{tab:SVCJ_pricing}
    \centering
    \small  
    \renewcommand{\arraystretch}{0.8} 
    \begin{tabular}{llcccccc}
        \toprule
        & & BS & Heston & SVCJ & ANN & NSDE & NJSDE \\
        \midrule
        In-sample & MAE & 1.1969 & 1.0194 & 0.6885 & 1.1346 & 0.8543 & 0.6127 \\
                  & MSE & 3.1573 & 2.4583 & 1.2674 & 2.5768 & 1.7692 & 0.7672 \\
        \midrule
        Out-of-sample & MAE & 1.3619 & 1.2059 & 0.9020 & 1.0572 & 0.9565 & 0.7698 \\
                      & MSE & 3.0744 & 2.4027 & 1.5696 & 2.0038 & 1.7394 & 0.9955 \\
        \bottomrule
    \end{tabular}
    \vspace{0.5em}
    
    \raggedright
    \footnotesize{\textit{Notes.} This table summarizes the in-sample and out-of-sample forecasting performance for synthetic options generated from the SVCJ model, using mean absolute error (MAE) and mean squared error (MSE).} 
\end{table}

Table \ref{tab:SVCJ_pricing} presents the overall performance across different models. As shown, the NJSDE model achieves the lowest in-sample and out-of-sample errors, suggesting its superior ability to capture jump dynamics in option pricing.
Among parametric models, the pricing accuracy improves with model complexity. The SVCJ model significantly outperforms the BS and Heston models, confirming the importance of jumps in both the asset price and volatility processes.
The ANN model, while achieving comparable performance to the Heston model, fails to surpass the SVCJ model. This result suggests that although neural networks can approximate non-linear dynamics, their effectiveness is constrained by data availability, particularly in capturing complex jump dynamics.

For hybrid models that integrate neural networks as parameter estimators, both the NSDE and NJSDE models exhibit superior performance compared to their respective parametric models, the Heston and SVCJ models, demonstrating the advantages of integrating the data-driven method with structural modeling frameworks. 
Notably, compared to the SVCJ model, the NSDE model demonstrates slightly lower predictive accuracy. Although \cite{wang2021option} does not provide a direct comparison between these two models, our findings indicate that the effectiveness of the NSDE model may be limited in conditions with pronounced jump dynamics and sparse training data. This highlights the need to improve the NSDE model by integrating a jump process, particularly when training data is limited. 

\begin{figure}[h]
    \caption{\textbf{Pricing Performances of the Competing Models across Different Moneyness and Maturities}}
    \label{fig:Errors in experiment1}
    \begin{subfigure}{0.495\textwidth}
        \centering
        \includegraphics[width=\textwidth]{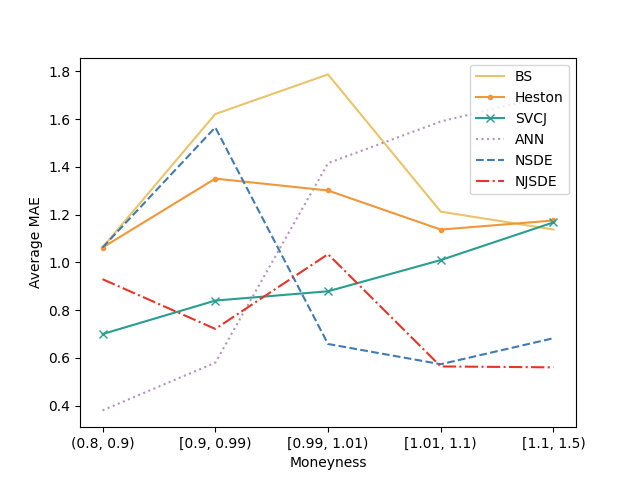}
        \caption{\small Out-of-Sample: All Maturities}
        \label{fig:E1_all}
    \end{subfigure}
    \hfill
    \begin{subfigure}{0.495\textwidth}
        \centering
        \includegraphics[width=\textwidth]{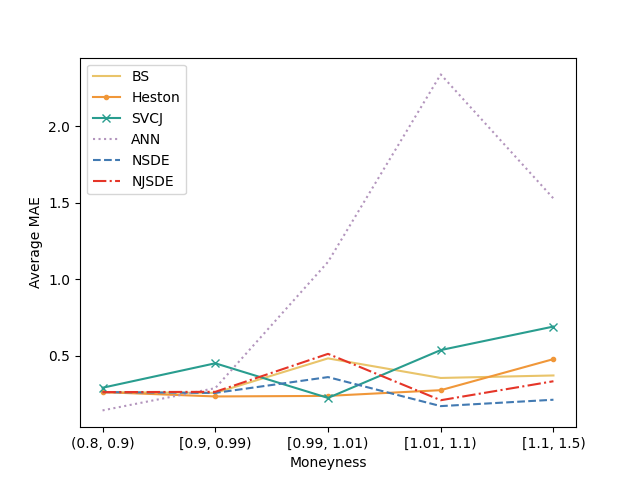}
        \caption{\small Out-of-Sample: Short-Term Maturity}
        \label{fig:E1_short}
    \end{subfigure}
    \hfill
    \begin{subfigure}{0.495\textwidth}
        \centering
        \includegraphics[width=\textwidth]{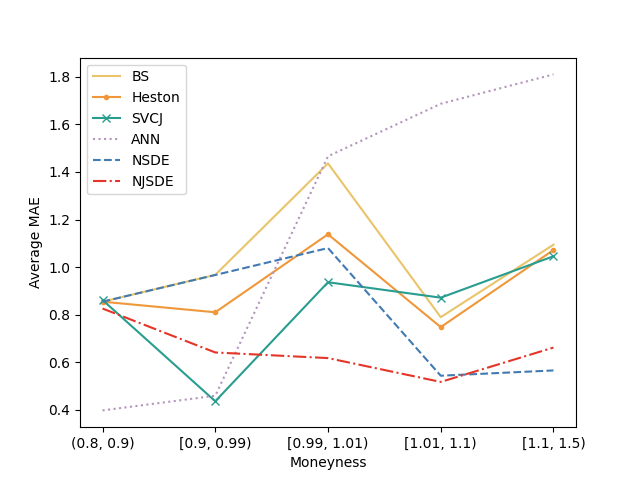}
        \caption{\small Out-of-Sample: Medium-Term Maturity}
        \label{fig:E1_medium}
    \end{subfigure}
    \hfill
    \begin{subfigure}{0.495\textwidth}
        \centering
        \includegraphics[width=\textwidth]{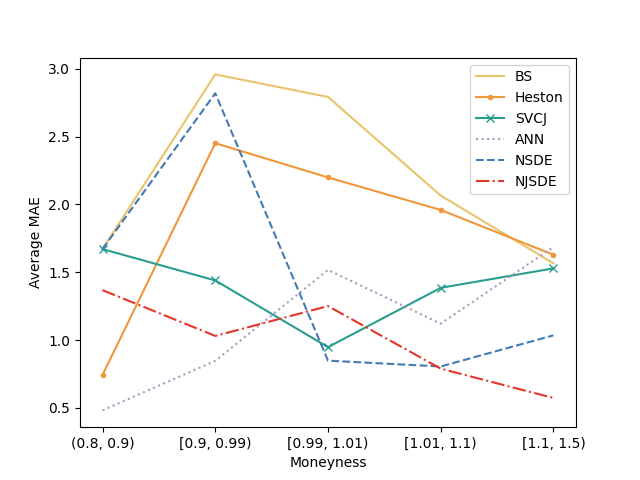}
        \caption{\small Out-of-Sample: Long-Term Maturity}
        \label{fig:E1_long}
    \end{subfigure}

    \smallskip
    \footnotesize\textit{Notes.} This figure shows the model’s out-of-sample performances across different Moneyness and maturity brackets on the synthetic options generated from SVCJ model, as measured by averages MAEs.
\end{figure}

Figure \ref{fig:Errors in experiment1} illustrates the error distribution across various maturity and moneyness levels. 
As shown in Figure \ref{fig:E1_all}, the NJSDE model consistently achieves the smallest or second smallest errors across all maturity and moneyness groups. While the ANN model exhibits the lowest errors in deep OTM, its performance deteriorates as the moneyness increases, particularly in the deep ITM. 
The advantage of the NJSDE model is relatively small for the short-term maturity shown in Figure \ref{fig:E1_short}. However, it increases with longer maturities, as illustrated in Figures \ref{fig:E1_medium} and \ref{fig:E1_long}.

\begin{table}[h] 
    \setlength{\tabcolsep}{8pt}  
    \caption{\textbf{DM Test of Different Models for Synthetic Options Generated from the SVCJ Model}}
    \label{tab:DM_1}
    \centering
    \small  
    \renewcommand{\arraystretch}{0.8} 
    \begin{tabular}{ c c c c c c }
        \toprule
        Model & Heston & SVCJ & ANN & NSDE & NJSDE \\
        \midrule
        BS & 
        3.94 (p$<$0.01) & 3.80 (p$<$0.01) & 2.56 (p$=$0.01) & 5.52 (p$<$0.01) &
        6.24 (p$<$0.01) \\
        Heston & - &
        2.65 (p$<$0.01) & 1.10 (p$=$0.27) & 3.26 (p$<$0.01) & 5.62 (p$<$0.01) \\
        SVCJ & - & - &
        1.48 (p$=$0.13) & 0.52 (p$=$0.60) & 2.30 (p$=$0.02) \\
        ANN & - & - & - &
        0.72 (p$=$0.46) & 4.32 (p$<$0.01) \\
        NSDE & - & - & - & - &
        2.95 (p$<$0.01) \\
        \bottomrule
    \end{tabular}
    \vspace{0.5em}
    
    \raggedright
    \footnotesize{\textit{Notes.} This table presents the DM values and p‐values for pairwise comparisons. A positive DM statistic indicates a preference for the column model over the row model.} 
\end{table}

Table \ref{tab:DM_1} presents the DM test results. 
The results show that the null hypothesis of equal predictive accuracy can be rejected at the 1\% significance level in most comparisons involving the NJSDE model, suggesting that it significantly outperforms the benchmark models.
Notably, the NJSDE model also significantly outperforms the SVCJ model, with a DM statistic of 2.30 (p = 0.02), which supports rejection of the null hypothesis at the 2\% significance level.
The DM value of the SVCJ model and the NSDE model is 0.52 with a p-value of 0.60, which supports the findings presented previously in Table \ref{tab:SVCJ_pricing}, indicating that while the NSDE model underperforms the SVCJ model, the difference is not statistically significant.
Lastly, there is no significant difference between the ANN model and the remaining models, except for the BS and NJSDE models.

\section{Empirical Analysis}\label{Real Data}

We further assess the model’s performance based on real market data. 
European-style S\&P 500 call options (SPX) data is obtained from OptionMetrics, covering the period from January 2, 2018 to December 30, 2022. The in-sample period spans from January 2, 2018, to December 31, 2021, while the out-of-sample period extends from January 3, 2022, to December 30, 2022. 
Additionally, we adopt zero-coupon yield curve from OptionMetrics and apply linear interpolation to align with each option’s maturity (\citeauthor{chung2011information}, \citeyear{chung2011information}).

Following \cite{ruf2022hedging}, we apply several filters to eliminate illiquid options.
Specifically, we discard observations that meet any of the following criteria: zero trading volume or open interest; bid price below 0.05 and ask price exceeding twice the bid price; time to expiration of less than one calendar day; moneyness outside the range of 0.80 to 1.50; violation of the lower boundary condition for European call option values.
After removing illiquid options, the final dataset consists of 591,284 SPX options across 1,258 trading days, with a daily average of approximately 470 options.

\begin{table}[h]
    \setlength{\tabcolsep}{8pt}  
    \begin{center}
    \caption{\textbf{Overview of S\&P 500 Option Data Summary}}
    \label{tab:sp500_summary}
    \small  
    \renewcommand{\arraystretch}{0.8} 
    \begin{tabular}{lcccccc}
        \toprule
        \textit{Moneyness} & (0.8, 0.9) & [0.9,0.99) & [0.99, 1.01) & [1.01, 1.1) & [1.1, 1.5) & Subtotal\\
        \midrule
        \multicolumn{7}{c}{\textbf{Panel A: Number of option contracts}} \\
        \midrule
        Short-term & $19,516$ & $113,590$ & $33,748$ & $60,667$ & $18,284$ & $245,805$ \\
        Medium-term & $33,919$ & $96,028$ & $25,614$ & $33,204$ & $14,173$ & $202,938$ \\
        Long-term & $40,176$ & $51,728$ & $14,276$ & $22,733$ & $13,628$ & $142,541$ \\
        Subtotal & $93,611$ & $261,346$ & $73,638$ & $116,604$ & $46,085$ & $591,284$ \\
        \midrule
        \multicolumn{7}{c}{\textbf{Panel B: Average option prices}} \\
        \midrule
        Short-term & $3.92$ & $19.88$ & $67.01$ & $172.77$ & $590.92$ & $105.29$ \\
        Medium-term & $15.21$ & $63.32$ & $140.90$ & $252.54$ & $670.82$ & $138.46$ \\
        Long-term & $69.04$ & $179.60$ & $290.88$ & $395.84$ & $756.39$ & $249.22$ \\
        Subtotal & $35.96$ & $67.45$ & $136.11$ & $238.98$ & $664.43$ & $151.37$ \\
        \bottomrule
    \end{tabular}
    \end{center}
    \vspace{0.5em}  
    \raggedright
    \footnotesize{\textit{Notes.} Moneyness is measured as $S/K$. We classify options into three maturity groups: short-term ($DTM$ $<$ 60 days), medium-term (60 $\leq$ $DTM$ $<$ 180 days), and long-term ($DTM$ $\geq$ 180 days).}
\end{table}

Table \ref{tab:sp500_summary} summarizes descriptive statistics of S\&P 500 index options, categorized according to days to maturity ($DTM$) and moneyness levels ($S/K$). 
Panel A reports the number of option contracts, showing that short-term options dominate the dataset, with a total of 245,805 contracts. Additionally, the dataset contains a disproportionately large number of OTM options relative to ITM options.
Average prices are shown in Panel B. Option prices exhibit a strong positive correlation with maturity, increasing from an average of 3.92 for short-term deep OTM options to 756.39 for long-term deep ITM options.
Moreover, for a given maturity, ITM options are priced significantly higher than OTM options.

\begin{figure}[h]
    \caption{\textbf{Monthly Dynamics of S\&P 500 Index Options}}
    \label{fig:sp500_summary}
    \begin{subfigure}{0.495\textwidth}
        \centering
        \includegraphics[width=\linewidth, height=5cm]{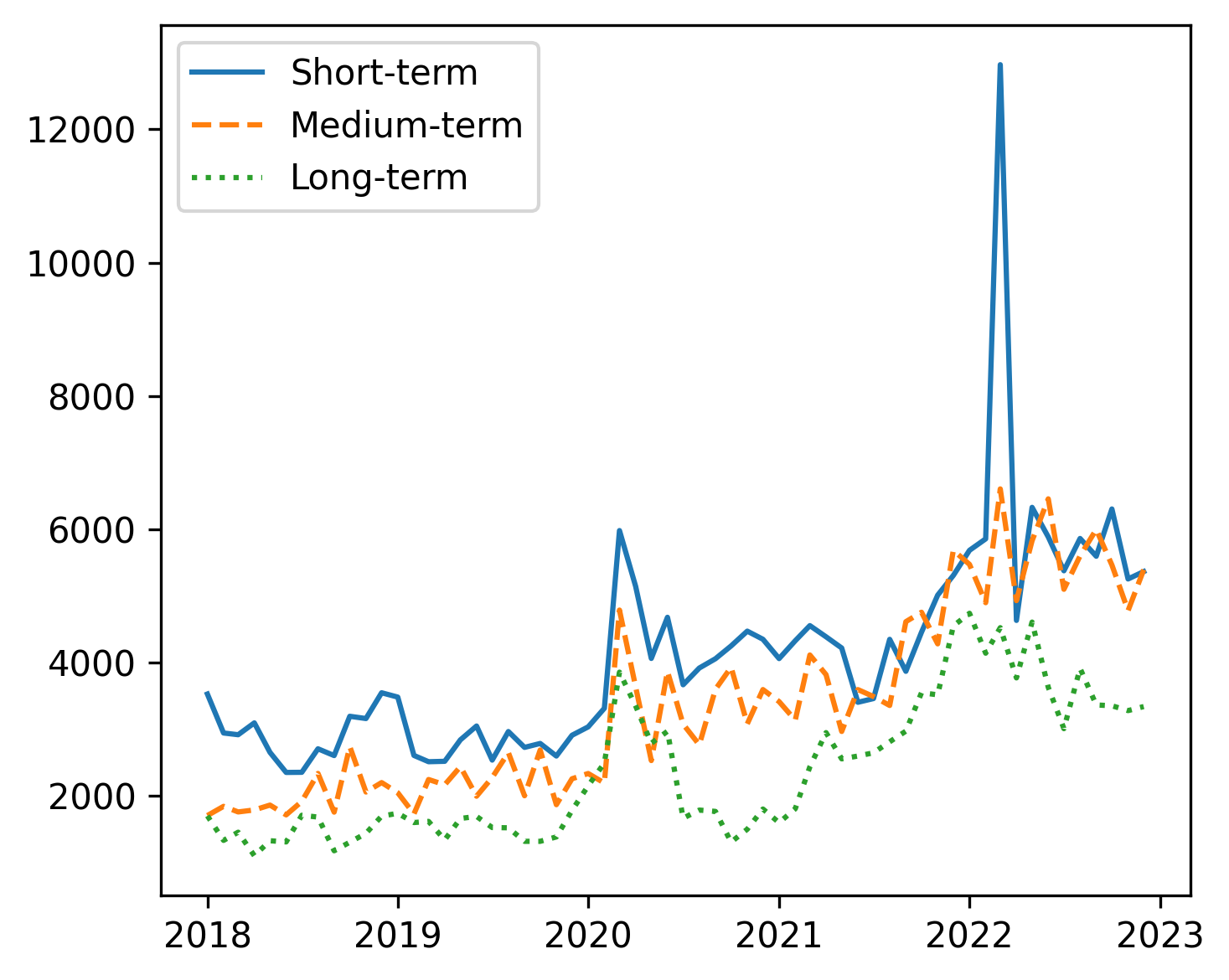} 
        \caption{\small Monthly Contract Counts across Maturity}
        \label{fig:sp500_summary_DTM}
    \end{subfigure}
    \hfill
    \begin{subfigure}{0.495\textwidth}
        \centering
        \includegraphics[width=\linewidth, height=5cm]{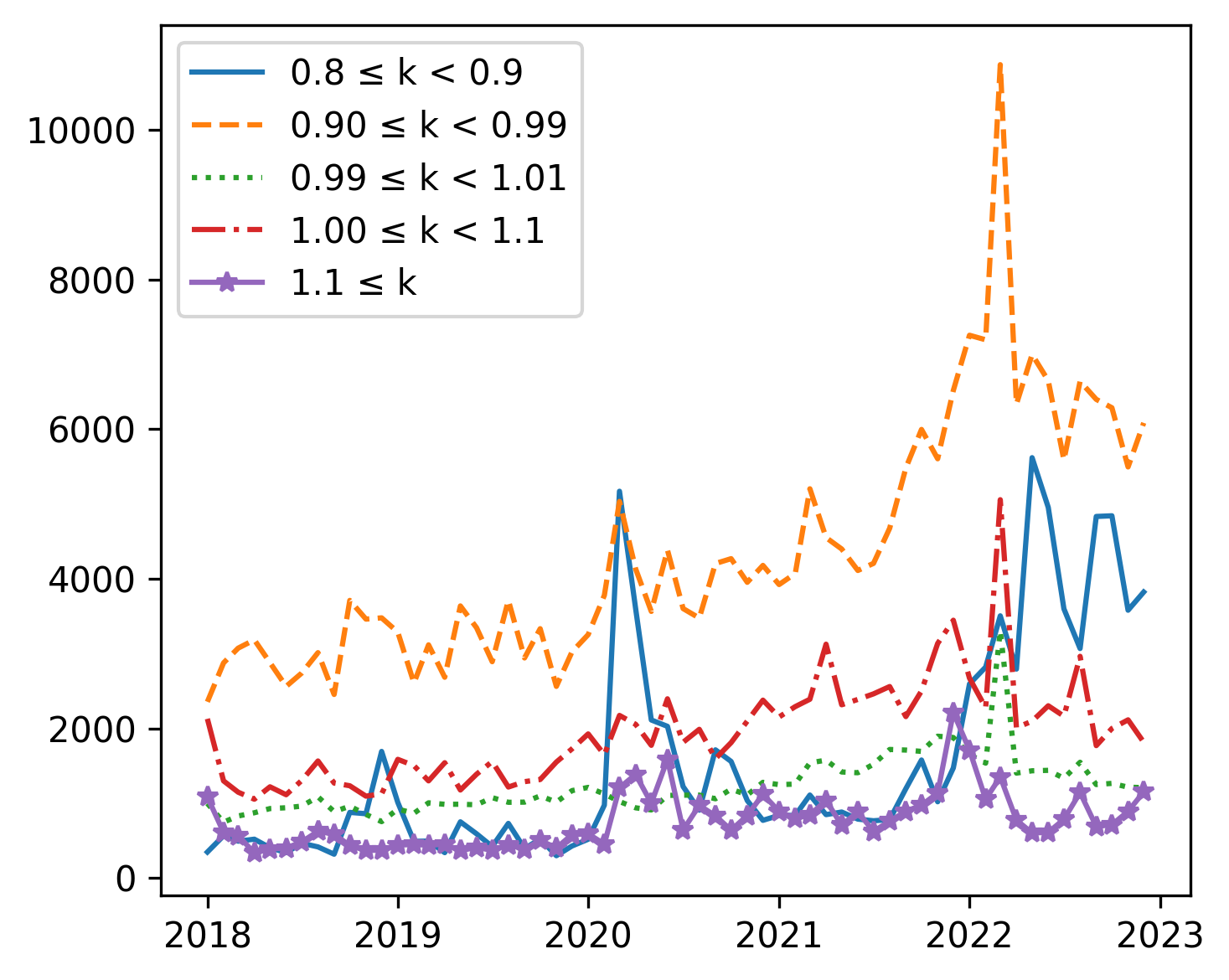}
        \caption{\small Monthly Contract Counts across Moneyness}
        \label{fig:sp500_summary_k}
    \end{subfigure}
    
    \smallskip
    \footnotesize\textit{Notes.} Panels (a) and (b) depict the monthly number of observed option contracts sorted by maturity and moneyness, respectively. Moneyness is measured by the spot-to-strike ratio ($S/K$), and \textit{DTM} refers to the number of calendar days remaining until expiration.
\end{figure}

Figure \ref{fig:sp500_summary} presents time series plots illustrating the trading dynamics of SPX options over time.
Figure \ref{fig:sp500_summary_DTM} depicts the temporal variation in the number of option contracts categorized by $DTM$. Trading volume is substantially concentrated in short-term options, likely driven by their superior liquidity and lower transaction costs. 
The spikes in trading volume are closely related to periods of market stress, as evidenced during COVID-19 in early 2020. In particular, short-term options experienced the most significant increase.
Figure \ref{fig:sp500_summary_k} displays time series patterns in the option volume, classified by distinct moneyness groups. OTM options are the most actively traded and deep OTM options have shown an increase in trading volume over time. In contrast, deep ITM options are relatively less traded, with trading volume remaining stable at approximately 1,000 contracts per day.

\begin{table}[h]
    \setlength{\tabcolsep}{8pt}  
    \caption{\textbf{Overall S\&P 500 Options Pricing Performance}}
    \label{tab:sp500_pricing}
    \centering
    \small  
    \renewcommand{\arraystretch}{0.8} 
    \begin{tabular}{llcccccc}
        \toprule
        & & BS & Heston & SVCJ & ANN & NSDE & NJSDE \\
        \midrule
        In-sample & MAE & 5.7030 & 4.8872 & 4.5126 & 3.1673 & 3.9246 & 2.9324 \\
                  & MSE & 49.3829 & 64.5714 & 64.1508 & 36.0628 & 54.9145 & 32.8247 \\
        \midrule
        Out-of-sample & MAE & 6.4308 & 4.4616 & 4.2944 & 4.0760 & 2.6794 & 2.4688 \\
                      & MSE & 59.8742 & 55.1507 & 46.3451 & 38.4611 & 15.6308 & 10.6443 \\
        \bottomrule
    \end{tabular}
    \vspace{0.5em}
    
    \raggedright
    \footnotesize{\textit{Notes.} This table summarizes the S\&P 500 options pricing errors, using mean absolute error (MAE) and mean squared error (MSE).} 
\end{table}

Table \ref{tab:sp500_pricing} summarizes the pricing performance in terms of MAE and MSE. We utilize the same evaluation metrics and benchmark models as in the simulation experiments.
The findings demonstrate that the NJSDE model outperforms other baseline models in forecasting performance.
Compared with the simulation results, two main differences are observed in the empirical findings. 
First, the NSDE model surpasses the SVCJ model in both in-sample and out-of-sample performance. This suggests that, given a sufficiently large dataset for training neural networks, a jump-free hybrid model can achieve lower pricing errors than a jump diffusion structural model, likely due to the enhanced approximation capabilities of neural networks.
Second, the ANN model achieves superior in-sample performance compared to the NSDE model, though its performance remains slightly inferior to that of the NJSDE model. This can potentially be explained by the neural network’s strong approximation ability in large datasets and the absence of a jump process in the NSDE model. 
Although the ANN model performs well in-sample, it exhibits significantly higher out-of-sample errors than both the NJSDE and NSDE models, indicating possible overfitting.
In conclusion, the results validate that jumps are present in S\&P 500 options data and highlight the importance of incorporating jumps into pricing models.
The NJSDE model, in particular, captures these jump-related features, demonstrating improved pricing accuracy.

\begin{figure}[h]
    \caption{\textbf{Pricing Performances of the Competing Models across Different Moneyness and Maturities}}
    \label{fig:Errors in experiment3}
    \begin{subfigure}{0.495\textwidth}
        \centering
        \includegraphics[width=\textwidth]{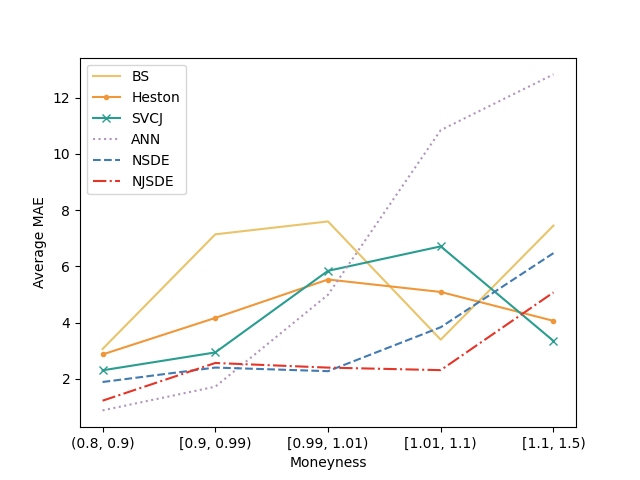}
        \caption{\small Out-of-Sample: All Maturities}
        \label{fig:E3_all}
    \end{subfigure}
    \hfill
    \begin{subfigure}{0.495\textwidth}
        \centering
        \includegraphics[width=\textwidth]{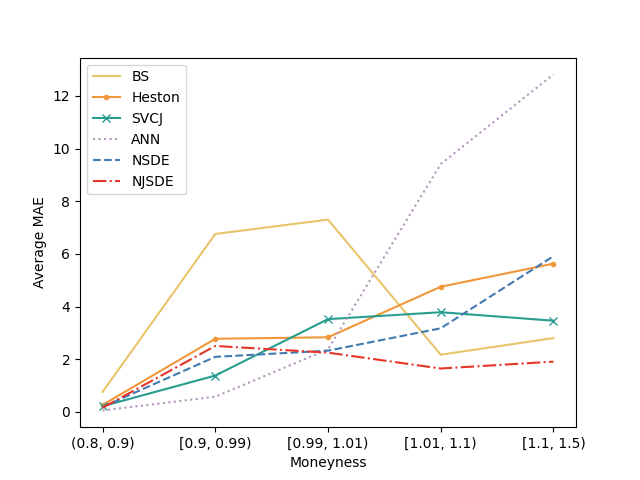}
        \caption{\small Out-of-Sample: Short-Term Maturity}
        \label{fig:E3_short}
    \end{subfigure}
    \hfill
    \begin{subfigure}{0.495\textwidth}
        \centering
        \includegraphics[width=\textwidth]{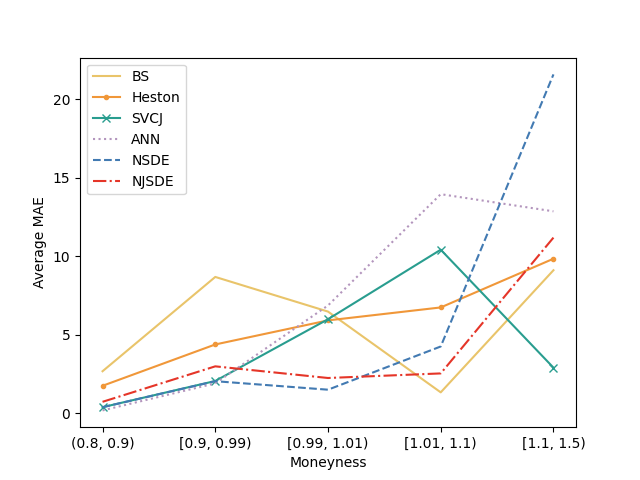}
        \caption{\small Out-of-Sample: Medium-Term Maturity}
        \label{fig:E3_medium}
    \end{subfigure}
    \hfill
    \begin{subfigure}{0.495\textwidth}
        \centering
        \includegraphics[width=\textwidth]{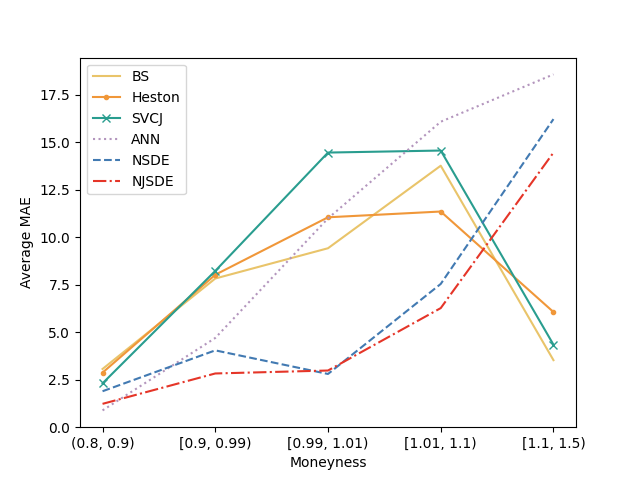}
        \caption{\small Out-of-Sample: Long-Term Maturity}
        \label{fig:E3_long}
    \end{subfigure}

    \smallskip
    \footnotesize\textit{Notes.} This figure shows the model’s out-of-sample performances across different Moneyness and maturity brackets on the S\&P 500 index options, as measured by averages MAEs.
\end{figure}

Figure \ref{fig:Errors in experiment3} provides a detailed evaluation by categorizing options according to moneyness and days to maturity ($DTM$). 
Overall, Figure \ref{fig:E3_all} indicates that the error distribution of the NJSDE model remains relatively stable across different moneyness bins. It consistently exhibits the lowest or second-lowest pricing errors among all models. However, for options with moneyness in the range [1.1, 1.5), the errors are noticeably larger, and this discrepancy becomes more pronounced as maturity increases.
Figure \ref{fig:E3_short} shows that the NSDE and NJSDE models outperform traditional models for short-term options, particularly those near the ATM, where they exhibit lower pricing errors. 
However, in Figures \ref{fig:E3_medium} and \ref{fig:E3_long}, the NSDE and NJSDE models exhibit higher MAE for ITM and OTM options, particularly those with long maturities and far from ATM. This may be due to the limited flexibility in deep ITM options, where parametric models tend to be less prone to misspecification.

\begin{table}[h] 
    \setlength{\tabcolsep}{8pt}  
    \caption{\textbf{DM Test of Different Models for S\&P 500 Options}}
    \label{tab:DM_3}
    \centering
    \small  
    \renewcommand{\arraystretch}{0.8} 
    \begin{tabular}{ c c c c c c }
        \toprule
        Model & Heston & SVCJ & ANN & NSDE & NJSDE \\
        \midrule
        BS & 
        0.51 (p$=0.60$) & 2.89 (p$<$0.01) & 4.49 (p$<$0.01) & 9.66 (p$<$0.01) &
        11.85 (p$<$0.01) \\
        Heston & - &
        1.08 (p$=0.27$) & 1.81 (p$=0.07$) & 4.30 (p$<$0.01) & 4.58 (p$<$0.01) \\
        SVCJ & - & - &
        1.17 (p$=$0.24) & 5.62 (p$<$0.01) & 6.98 (p$<$0.01) \\
         ANN & - & - & - &
        6.21 (p$<$0.01) & 8.32 (p$<$0.01) \\
        NSDE & - & - & - & - &
        2.64 (p$<$0.01) \\
        \bottomrule
    \end{tabular}
    \vspace{0.5em}
    
    \raggedright
    \footnotesize{\textit{Notes.} This table presents the DM values and p‐values for pairwise comparisons. A positive DM statistic indicates a preference for the column model over the row model.} 
\end{table}

Table \ref{tab:DM_3} presents the DM values for S\&P 500 options. This evidence is consistent with previous observations. Most importantly, the final column reports that the NJSDE model has consistently and significantly lower forecast errors compared to other models, demonstrating its robustness in S\&P 500 option pricing. Compared to the SVCJ model, the NSDE model demonstrates statistically superior performance, with the difference being significant at the 1\% level. 
Similarly, this discrepancy between the two models can be explained by the larger dataset available in the real data experiment compared to the simulated data. These findings suggest that data availability influences the performance of hybrid models and that incorporating structural constraints can mitigate sensitivity to data volume. 
Additionally, there is no statistically significant difference in forecasting performance between the ANN model and the Heston or SVCJ models at conventional significance levels.

\section{Conclusion}\label{conclusion}
In this paper, we establish a hybrid model for integrating jump theoretical structures with neural networks. As jump risk is a key determinant in option pricing, we construct a gradient learnable jump process that employs the Gumbel-Softmax method to accommodate discontinuities in the jump process and stochastic dependence on the trainable parameter. 
By integrating these two influential models, neural networks and jump diffusion model, the proposed hybrid model achieves both economic interpretability and strong approximation capabilities. 
Empirical results demonstrate that the proposed model provides superior predictive accuracy compared to the benchmark models under jump conditions. In a jump-free setting, it exhibits comparable performance to the NSDE model, demonstrating adaptability in different market periods.

\bibliographystyle{elsarticle-harv} 
\bibliography{mybibfile} 

\begin{thebibliography}{40}
\expandafter\ifx\csname natexlab\endcsname\relax\def\natexlab#1{#1}\fi
\providecommand{\url}[1]{\texttt{#1}}
\providecommand{\href}[2]{#2}
\providecommand{\path}[1]{#1}
\providecommand{\DOIprefix}{doi:}
\providecommand{\ArXivprefix}{arXiv:}
\providecommand{\URLprefix}{URL: }
\providecommand{\Pubmedprefix}{pmid:}
\providecommand{\doi}[1]{\href{http://dx.doi.org/#1}{\path{#1}}}
\providecommand{\Pubmed}[1]{\href{pmid:#1}{\path{#1}}}
\providecommand{\bibinfo}[2]{#2}
\ifx\xfnm\relax \def\xfnm[#1]{\unskip,\space#1}\fi
\bibitem[{Amilon(2003)}]{amilon2003neural}
\bibinfo{author}{Amilon, H.}, \bibinfo{year}{2003}.
\newblock \bibinfo{title}{A neural network versus {Black--Scholes}: {A}
  comparison of pricing and hedging performances}.
\newblock \bibinfo{journal}{Journal of Forecasting} \bibinfo{volume}{22},
  \bibinfo{pages}{317--335}.
\bibitem[{Andreou et~al.(2008)Andreou, Charalambous and
  Martzoukos}]{andreou2008pricing}
\bibinfo{author}{Andreou, P.C.}, \bibinfo{author}{Charalambous, C.},
  \bibinfo{author}{Martzoukos, S.H.}, \bibinfo{year}{2008}.
\newblock \bibinfo{title}{Pricing and trading {European} options by combining
  artificial neural networks and parametric models with implied parameters}.
\newblock \bibinfo{journal}{European Journal of Operational Research}
  \bibinfo{volume}{185}, \bibinfo{pages}{1415--1433}.
\bibitem[{Andreou et~al.(2010)Andreou, Charalambous and
  Martzoukos}]{andreou2010generalized}
\bibinfo{author}{Andreou, P.C.}, \bibinfo{author}{Charalambous, C.},
  \bibinfo{author}{Martzoukos, S.H.}, \bibinfo{year}{2010}.
\newblock \bibinfo{title}{Generalized parameter functions for option pricing}.
\newblock \bibinfo{journal}{Journal of Banking \& Finance}
  \bibinfo{volume}{34}, \bibinfo{pages}{633--646}.
\bibitem[{Bates(1996)}]{bates1996jumps}
\bibinfo{author}{Bates, D.S.}, \bibinfo{year}{1996}.
\newblock \bibinfo{title}{Jumps and stochastic volatility: {Exchange} rate
  processes implicit in deutsche mark options}.
\newblock \bibinfo{journal}{The Review of Financial Studies}
  \bibinfo{volume}{9}, \bibinfo{pages}{69--107}.
\bibitem[{Bates(2003)}]{bates2003empirical}
\bibinfo{author}{Bates, D.S.}, \bibinfo{year}{2003}.
\newblock \bibinfo{title}{Empirical option pricing: {A} retrospection}.
\newblock \bibinfo{journal}{Journal of Econometrics} \bibinfo{volume}{116},
  \bibinfo{pages}{387--404}.
\bibitem[{Black and Scholes(1973)}]{black1973pricing}
\bibinfo{author}{Black, F.}, \bibinfo{author}{Scholes, M.},
  \bibinfo{year}{1973}.
\newblock \bibinfo{title}{The pricing of options and corporate liabilities}.
\newblock \bibinfo{journal}{Journal of Political Economy} \bibinfo{volume}{81},
  \bibinfo{pages}{637--654}.
\bibitem[{Cao et~al.(2021)Cao, Liu and Zhai}]{cao2021option}
\bibinfo{author}{Cao, Y.}, \bibinfo{author}{Liu, X.}, \bibinfo{author}{Zhai,
  J.}, \bibinfo{year}{2021}.
\newblock \bibinfo{title}{Option valuation under no-arbitrage constraints with
  neural networks}.
\newblock \bibinfo{journal}{European Journal of Operational Research}
  \bibinfo{volume}{293}, \bibinfo{pages}{361--374}.
\bibitem[{Chen et~al.(2025)Chen, Didisheim and Scheidegger}]{chen2023deep}
\bibinfo{author}{Chen, H.}, \bibinfo{author}{Didisheim, A.},
  \bibinfo{author}{Scheidegger, S.}, \bibinfo{year}{2025}.
\newblock \bibinfo{title}{Deep surrogates for finance: {With} an application to
  option pricing}.
\newblock \bibinfo{journal}{Journal of Financial Economics.}
  \bibinfo{note}{Forthcoming}.
\bibitem[{Chen et~al.(2018)Chen, Rubanova, Bettencourt and
  Duvenaud}]{chen2018neural}
\bibinfo{author}{Chen, R.T.}, \bibinfo{author}{Rubanova, Y.},
  \bibinfo{author}{Bettencourt, J.}, \bibinfo{author}{Duvenaud, D.K.},
  \bibinfo{year}{2018}.
\newblock \bibinfo{title}{Neural ordinary differential equations}.
\newblock \bibinfo{journal}{Advances in Neural Information Processing Systems
  (NeurIPS)} \bibinfo{volume}{31}.
\bibitem[{Chung et~al.(2011)Chung, Tsai, Wang and Weng}]{chung2011information}
\bibinfo{author}{Chung, S.L.}, \bibinfo{author}{Tsai, W.C.},
  \bibinfo{author}{Wang, Y.H.}, \bibinfo{author}{Weng, P.S.},
  \bibinfo{year}{2011}.
\newblock \bibinfo{title}{The information content of the {S\&P} 500 index and
  {VIX} options on the dynamics of the {S\&P} 500 index}.
\newblock \bibinfo{journal}{Journal of Futures Markets} \bibinfo{volume}{31},
  \bibinfo{pages}{1170--1201}.
\bibitem[{Cox and Ross(1976)}]{cox1976valuation}
\bibinfo{author}{Cox, J.C.}, \bibinfo{author}{Ross, S.A.},
  \bibinfo{year}{1976}.
\newblock \bibinfo{title}{The valuation of options for alternative stochastic
  processes}.
\newblock \bibinfo{journal}{Journal of Financial Economics}
  \bibinfo{volume}{3}, \bibinfo{pages}{145--166}.
\bibitem[{Cummins and Esposito(2025)}]{cummins2025appraising}
\bibinfo{author}{Cummins, M.}, \bibinfo{author}{Esposito, F.},
  \bibinfo{year}{2025}.
\newblock \bibinfo{title}{Appraising model complexity in option pricing}.
\newblock \bibinfo{journal}{Journal of Futures Markets} \bibinfo{volume}{45},
  \bibinfo{pages}{455--472}.
\bibitem[{Das and Padhy(2017)}]{das2017new}
\bibinfo{author}{Das, S.P.}, \bibinfo{author}{Padhy, S.}, \bibinfo{year}{2017}.
\newblock \bibinfo{title}{A new hybrid parametric and machine learning model
  with homogeneity hint for {European-style} index option pricing}.
\newblock \bibinfo{journal}{Neural Computing and Applications}
  \bibinfo{volume}{28}, \bibinfo{pages}{4061--4077}.
\bibitem[{Diebold and Mariano(2002)}]{diebold2002comparing}
\bibinfo{author}{Diebold, F.X.}, \bibinfo{author}{Mariano, R.S.},
  \bibinfo{year}{2002}.
\newblock \bibinfo{title}{Comparing predictive accuracy}.
\newblock \bibinfo{journal}{Journal of Business \& Economic Statistics}
  \bibinfo{volume}{20}, \bibinfo{pages}{134--144}.
\bibitem[{Duffie et~al.(2000)Duffie, Pan and Singleton}]{duffie2000transform}
\bibinfo{author}{Duffie, D.}, \bibinfo{author}{Pan, J.},
  \bibinfo{author}{Singleton, K.}, \bibinfo{year}{2000}.
\newblock \bibinfo{title}{Transform analysis and asset pricing for affine
  jump-diffusions}.
\newblock \bibinfo{journal}{Econometrica} \bibinfo{volume}{68},
  \bibinfo{pages}{1343--1376}.
\bibitem[{Dupire et~al.(1994)}]{dupire1994pricing}
\bibinfo{author}{Dupire, B.}, et~al., \bibinfo{year}{1994}.
\newblock \bibinfo{title}{Pricing with a smile}.
\newblock \bibinfo{journal}{Risk} \bibinfo{volume}{7}, \bibinfo{pages}{18--20}.
\bibitem[{E(2017)}]{weinan2017proposal}
\bibinfo{author}{E, W.}, \bibinfo{year}{2017}.
\newblock \bibinfo{title}{A proposal on machine learning via dynamical
  systems}.
\newblock \bibinfo{journal}{Communications in Mathematics and Statistics}
  \bibinfo{volume}{1}, \bibinfo{pages}{1--11}.
\bibitem[{Eraker et~al.(2003)Eraker, Johannes and Polson}]{eraker2003impact}
\bibinfo{author}{Eraker, B.}, \bibinfo{author}{Johannes, M.},
  \bibinfo{author}{Polson, N.}, \bibinfo{year}{2003}.
\newblock \bibinfo{title}{The impact of jumps in volatility and returns}.
\newblock \bibinfo{journal}{The Journal of Finance} \bibinfo{volume}{58},
  \bibinfo{pages}{1269--1300}.
\bibitem[{Gumbel(1954)}]{gumbel1954statistical}
\bibinfo{author}{Gumbel, E.J.}, \bibinfo{year}{1954}.
\newblock \bibinfo{title}{Statistical theory of extreme values and some
  practical applications: {A} series of lectures}.
\newblock \bibinfo{publisher}{US Government Printing Office}.
\bibitem[{Halskov(2023)}]{halskov2023deep}
\bibinfo{author}{Halskov, K.}, \bibinfo{year}{2023}.
\newblock \bibinfo{title}{A deep structural model for empirical asset pricing}.
\newblock \bibinfo{note}{Working Paper}.
\bibitem[{Heston(1993)}]{heston1993closed}
\bibinfo{author}{Heston, S.L.}, \bibinfo{year}{1993}.
\newblock \bibinfo{title}{A closed-form solution for options with stochastic
  volatility with applications to bond and currency options}.
\newblock \bibinfo{journal}{The Review of Financial Studies}
  \bibinfo{volume}{6}, \bibinfo{pages}{327--343}.
\bibitem[{Hornik et~al.(1989)Hornik, Stinchcombe and
  White}]{hornik1989multilayer}
\bibinfo{author}{Hornik, K.}, \bibinfo{author}{Stinchcombe, M.},
  \bibinfo{author}{White, H.}, \bibinfo{year}{1989}.
\newblock \bibinfo{title}{Multilayer feedforward networks are universal
  approximators}.
\newblock \bibinfo{journal}{Neural Networks} \bibinfo{volume}{2},
  \bibinfo{pages}{359--366}.
\bibitem[{Hull and White(1987)}]{hull1987pricing}
\bibinfo{author}{Hull, J.}, \bibinfo{author}{White, A.}, \bibinfo{year}{1987}.
\newblock \bibinfo{title}{The pricing of options on assets with stochastic
  volatilities}.
\newblock \bibinfo{journal}{The Journal of Finance} \bibinfo{volume}{42},
  \bibinfo{pages}{281--300}.
\bibitem[{Hutchinson et~al.(1994)Hutchinson, Lo and
  Poggio}]{hutchinson1994nonparametric}
\bibinfo{author}{Hutchinson, J.M.}, \bibinfo{author}{Lo, A.W.},
  \bibinfo{author}{Poggio, T.}, \bibinfo{year}{1994}.
\newblock \bibinfo{title}{A nonparametric approach to pricing and hedging
  derivative securities via learning networks}.
\newblock \bibinfo{journal}{The Journal of Finance} \bibinfo{volume}{49},
  \bibinfo{pages}{851--889}.
\bibitem[{Jang et~al.(2017)Jang, Gu and Poole}]{jang2017categorical}
\bibinfo{author}{Jang, E.}, \bibinfo{author}{Gu, S.}, \bibinfo{author}{Poole,
  B.}, \bibinfo{year}{2017}.
\newblock \bibinfo{title}{Categorical reparameterization with
  {Gumbel-Softmax}}.
\newblock \bibinfo{journal}{International Conference on Learning
  Representations (ICLR)} .
\bibitem[{Jia and Benson(2019)}]{jia2019neural}
\bibinfo{author}{Jia, J.}, \bibinfo{author}{Benson, A.R.},
  \bibinfo{year}{2019}.
\newblock \bibinfo{title}{Neural jump stochastic differential equations}.
\newblock \bibinfo{journal}{Advances in Neural Information Processing Systems
  (NeurIPS)} \bibinfo{volume}{32}.
\bibitem[{Khoo et~al.(2021)Khoo, Lu and Ying}]{khoo2021solving}
\bibinfo{author}{Khoo, Y.}, \bibinfo{author}{Lu, J.}, \bibinfo{author}{Ying,
  L.}, \bibinfo{year}{2021}.
\newblock \bibinfo{title}{Solving parametric {PDE} problems with artificial
  neural networks}.
\newblock \bibinfo{journal}{European Journal of Applied Mathematics}
  \bibinfo{volume}{32}, \bibinfo{pages}{421--435}.
\bibitem[{Kidger et~al.(2020)Kidger, Morrill, Foster and
  Lyons}]{kidger2020neural}
\bibinfo{author}{Kidger, P.}, \bibinfo{author}{Morrill, J.},
  \bibinfo{author}{Foster, J.}, \bibinfo{author}{Lyons, T.},
  \bibinfo{year}{2020}.
\newblock \bibinfo{title}{Neural controlled differential equations for
  irregular time series}.
\newblock \bibinfo{journal}{Advances in Neural Information Processing Systems
  (NeurIPS)} \bibinfo{volume}{33}, \bibinfo{pages}{6696--6707}.
\bibitem[{Kim(2021)}]{kim2021portfolio}
\bibinfo{author}{Kim, S.}, \bibinfo{year}{2021}.
\newblock \bibinfo{title}{Portfolio of volatility smiles versus volatility
  surface: {Implications} for pricing and hedging options}.
\newblock \bibinfo{journal}{Journal of Futures Markets} \bibinfo{volume}{41},
  \bibinfo{pages}{1154--1176}.
\bibitem[{Kou(2002)}]{kou2002jump}
\bibinfo{author}{Kou, S.G.}, \bibinfo{year}{2002}.
\newblock \bibinfo{title}{A jump-diffusion model for option pricing}.
\newblock \bibinfo{journal}{Management Science} \bibinfo{volume}{48},
  \bibinfo{pages}{1086--1101}.
\bibitem[{Li et~al.(2020)Li, Wong, Chen and Duvenaud}]{li2020scalable}
\bibinfo{author}{Li, X.}, \bibinfo{author}{Wong, T.K.L.},
  \bibinfo{author}{Chen, R.T.}, \bibinfo{author}{Duvenaud, D.},
  \bibinfo{year}{2020}.
\newblock \bibinfo{title}{Scalable gradients for stochastic differential
  equations}.
\newblock \bibinfo{journal}{International Conference on Artificial Intelligence
  and Statistics (AISTATS)} , \bibinfo{pages}{3870--3882}.
\bibitem[{Liu et~al.(2019)Liu, Oosterlee and Bohte}]{liu2019pricing}
\bibinfo{author}{Liu, S.}, \bibinfo{author}{Oosterlee, C.W.},
  \bibinfo{author}{Bohte, S.M.}, \bibinfo{year}{2019}.
\newblock \bibinfo{title}{Pricing options and computing implied volatilities
  using neural networks}.
\newblock \bibinfo{journal}{Risk} \bibinfo{volume}{7}, \bibinfo{pages}{16}.
\bibitem[{Ludwig(2015)}]{ludwig2015robust}
\bibinfo{author}{Ludwig, M.}, \bibinfo{year}{2015}.
\newblock \bibinfo{title}{Robust estimation of shape-constrained state price
  density surfaces}.
\newblock \bibinfo{journal}{The Journal of Derivatives} \bibinfo{volume}{22},
  \bibinfo{pages}{56--72}.
\bibitem[{Ma et~al.(2023)Ma, Wu and Wu}]{maneural}
\bibinfo{author}{Ma, J.}, \bibinfo{author}{Wu, X.}, \bibinfo{author}{Wu, H.},
  \bibinfo{year}{2023}.
\newblock \bibinfo{title}{A neural network rough volatility model}.
\newblock \bibinfo{note}{Working Paper}.
\bibitem[{Ma et~al.(2021)Ma, Dixit, Innes, Guo and
  Rackauckas}]{ma2021comparison}
\bibinfo{author}{Ma, Y.}, \bibinfo{author}{Dixit, V.}, \bibinfo{author}{Innes,
  M.J.}, \bibinfo{author}{Guo, X.}, \bibinfo{author}{Rackauckas, C.},
  \bibinfo{year}{2021}.
\newblock \bibinfo{title}{A comparison of automatic differentiation and
  continuous sensitivity analysis for derivatives of differential equation
  solutions}.
\newblock \bibinfo{journal}{2021 IEEE High Performance Extreme Computing
  Conference (HPEC)} , \bibinfo{pages}{1--9}.
\bibitem[{Merton(1976)}]{merton1976option}
\bibinfo{author}{Merton, R.C.}, \bibinfo{year}{1976}.
\newblock \bibinfo{title}{Option pricing when underlying stock returns are
  discontinuous}.
\newblock \bibinfo{journal}{Journal of Financial Economics}
  \bibinfo{volume}{3}, \bibinfo{pages}{125--144}.
\bibitem[{Ruf and Wang(2020)}]{ruf2020neural}
\bibinfo{author}{Ruf, J.}, \bibinfo{author}{Wang, W.}, \bibinfo{year}{2020}.
\newblock \bibinfo{title}{Neural networks for option pricing and hedging: {A}
  literature review}.
\newblock \bibinfo{journal}{Journal of Computational Finance}
  \bibinfo{volume}{24}, \bibinfo{pages}{1--46}.
\bibitem[{Ruf and Wang(2022)}]{ruf2022hedging}
\bibinfo{author}{Ruf, J.}, \bibinfo{author}{Wang, W.}, \bibinfo{year}{2022}.
\newblock \bibinfo{title}{Hedging with linear regressions and neural networks}.
\newblock \bibinfo{journal}{Journal of Business \& Economic Statistics}
  \bibinfo{volume}{40}, \bibinfo{pages}{1442--1454}.
\bibitem[{Shvimer and Zhu(2024)}]{shvimer2024pricing}
\bibinfo{author}{Shvimer, Y.}, \bibinfo{author}{Zhu, S.P.},
  \bibinfo{year}{2024}.
\newblock \bibinfo{title}{Pricing options with a new hybrid neural network
  model}.
\newblock \bibinfo{journal}{Expert Systems with Applications}
  \bibinfo{volume}{251}, \bibinfo{pages}{123979}.
\bibitem[{Wang and Hong(2021)}]{wang2021option}
\bibinfo{author}{Wang, S.}, \bibinfo{author}{Hong, L.J.}, \bibinfo{year}{2021}.
\newblock \bibinfo{title}{Option pricing by neural stochastic differential
  equations: {A} simulation-optimization approach}.
\newblock \bibinfo{journal}{2021 Winter Simulation Conference (WSC)} ,
  \bibinfo{pages}{1--11}.

\end{thebibliography}


\end{document}